\documentclass[epj]{webofc}
\usepackage[varg]{txfonts,epsfig}   
\wocname{EPJ Web of Conferences}
\woctitle{CONF12}
\begin{document}
\selectlanguage{english}
\title{Conditions for the existence of stable strange quark matter}
%
%

\author{N.A. Dondi\inst{1}\fnsep\thanks{\email{nicolaandrea.dondi@student.unife.it}} \and
        A. Drago\inst{1} \and
        G. Pagliara\inst{1}
}

\institute{Dip.~di Fisica e Scienze della Terra dell'Universit\`a di Ferrara and INFN
Sez.~di
Ferrara, Via Saragat 1, I-44100 Ferrara, Italy
}

\abstract{We discuss the possible existence of absolutely stable strange quark matter within
  three different types of chiral models. We will show that confinement plays
  a crucial role in determining the conditions for the Bodmer-Witten hypothesis to hold true. We discuss
  also which are the phenomenological signatures, related to measurements of masses and radii of compact stars,
  which would prove the existence of strange quark stars.}
\maketitle
\section{Introduction}
\label{intro}
The so called Bodmer-Witten hypothesis on the absolute stability of
strange quark matter (see \cite{Bodmer:1971we,Witten:1984rs} and \cite{Weber:2004kj} for a review) still
represents an open issue of nuclear and particle physics and its
validity would have numerous important phenomenological implications
in astrophysics (strange quark stars and strangelets in cosmic rays)
and, possibly, also in cosmology (strangelets as a component of
baryonic dark matter). Verifying this hypothesis in terrestrial
experiments is however very difficult because of the large net
strangeness carried by this form of strongly interacting matter. On
the other hand, theoretical calculations of the mass and density of
strange quark matter droplets and of the thermodynamic properties of
bulk strange quark matter are affected by large uncertainties due to
the intrinsic difficulty of solving QCD at finite (but not
asymptotically large) baryon densities. In this short contribution we
will analyze two aspects related to the possibility of the existence
of strange quark matter. First, we will show that this hypothesis can
be fulfilled also in a class of quark models which feature not only
confinement (as the MIT bag model) but also chiral
symmetry. Specifically, we will show results obtained within the
so-called chiral chromodielectric model \cite{Pirner:1991im}.

The second aspect concerns the astrophysical measurements of masses
and radii of compact stars. Precise radii measurements (which will be
feasible in the near future) could indeed unambiguously check the
validity of the Bodmer-Witten hypothesis.
Let us define $R_{1.4}$ as the radius of a $1.4 M_{\odot}$ compact star.
As found in \cite{Kurkela:2014vha,Alford:2015dpa} it is not possible
to have a unique family of compact stars satisfying at the same time
the condition $R_{1.4} \lesssim 11$km and $M_{max} \leq 2M_{\odot}$.
This conclusion has been confirmed also by other recent analyses \cite{lattimer-talk}. 
If future measurements will confirm the existence of stars
for which  $R_{1.4} \lesssim 11$km \cite{Ozel:2015fia}, then one can
conclude that strange quark stars co-exist with hadronic
stars, as proposed in
Refs.\cite{Drago:2013fsa,Drago:2015cea,Drago:2015dea}
In this two families scenario compact and light stars are hadronic stars whereas
large and massive stars are strange quark stars.

Since the $\sim 2M_{\odot}$ compact stars are, in this scenario,
interpreted as strange quark stars that implies that the density of
bound strange quark matter (i.e. the density corresponding to the
minimum of the energy per baryon) is between 1-2 times nuclear matter
saturation density. 
It is important to remark that although the absolute minimum
of the energy per baryon of nuclear matter is located
at a density similar to the density of strange quark matter
they differ concerning the strangeness content.
This implies that a direct transition from the metastable nucleonic
state to the ground state of quark matter is suppressed because it requires
multiple weak decays \cite{Witten:1984rs}.

\section{Quark matter equation of state}
\label{sec-1}

\begin{figure}[ptb]
\vskip 0.5cm
\begin{centering}
\epsfig{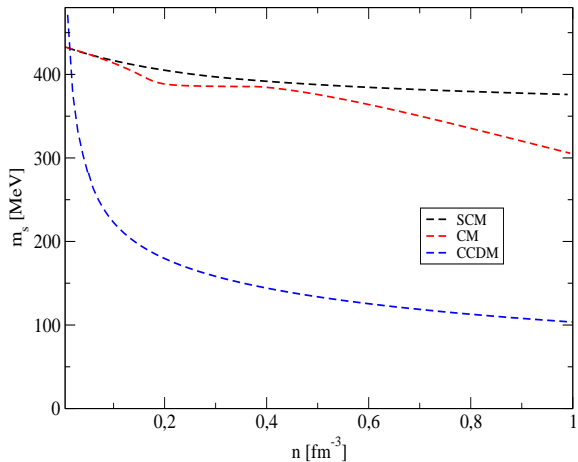}
\caption{Baryon density dependence of the strange quark mass in the three chiral models considered in this work.}
\end{centering}
\end{figure}
It is a fact that ab-initio QCD calculations in the regime of high baryon density
are still unfeasible and therefore to describe the matter of compact stars one has to rely on quark/hadronic
models which capture at least some of the features of QCD.
When exploring the conditions for the Bodmer-Witten hypothesis to hold true,
the MIT bag model (which implements in a simple way QCD confinement)
turns out to have a significant freedom in the choice of parameters
to allow for the existence of stable strange quark matter, see the seminal paper \cite{Farhi:1984qu}.
In particular, the Bodmer-Witten hypothesis is satisfied when small values
of the strange quark current mass are adopted.
On the other hand, in chiral models, such as the NJL model \cite{Buballa:1998pr}
this possibility seems to be ruled out, see also the recent work of Ref.~\cite{Klahn:2015mfa}.
Notice that while the MIT bag model, in its simplest version,
features confinement but not chiral symmetry, the opposite is true
in (simple) chiral models. It is therefore clear that the Bodmer-Witten
hypothesis could be fulfilled if a subtle interplay between
confinement and spontaneous chiral symmetry breaking exists. 
We consider here different kinds of SU(3) chiral models in which quarks
interact through meson exchange. What turns out to be the most important
microscopic quantity for the validity of the Bodmer-Witten hypothesis
is the density dependence of the strange quark mass which in turn
is determined by the quark-meson couplings.
To obtain a minimum of the energy per baryon lower than the one of nuclear matter (in order for strange quark matter to be absolutely stable),
it is necessary that the strange quark mass decreases significantly at densities close to nuclear saturation density.

\begin{figure}[ptb]
\vskip 0.5cm
\begin{centering}
\epsfig{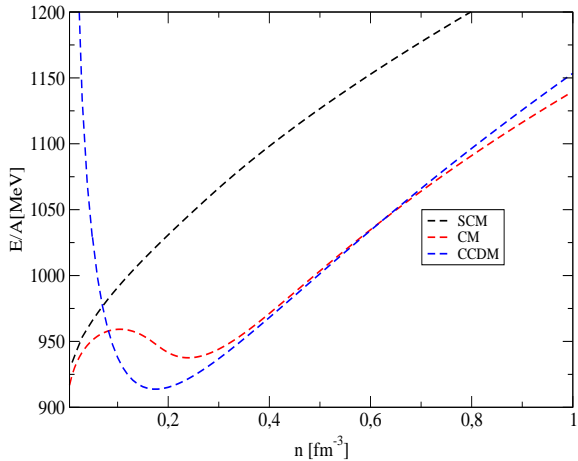}
\caption{Energy per baryon as a function of the baryon density for the three chiral models considered in this work.}
\end{centering}
\end{figure}

We analyze the density dependence of the strange quark mass by using three different chiral models for
beta-stable and charge neutral quark matter within the Hartree approximation:
\begin{itemize}
\item The basic version of the SU(3) Chiral Model (CM) which describes deconfined quarks
  interacting through scalar and vector mesons. 
\item The Scaled Chiral Model (SCM) which additionally incorporates the
  anomalous breaking of the scale invariance through the action of the
  dilaton scalar field.
\item The Chiral Chromodielectric model (CCDM) which allows to simulate quark confinement
  through the action of the color dielectric field.
\end{itemize}
Among the three models, only the CCDM features dynamical confinement. Indeed in this model
the effect of the dielectric field is to make the quark masses divergent in vacuum.

\begin{figure}[ptb]
\vskip 0.5cm
\begin{centering}
\epsfig{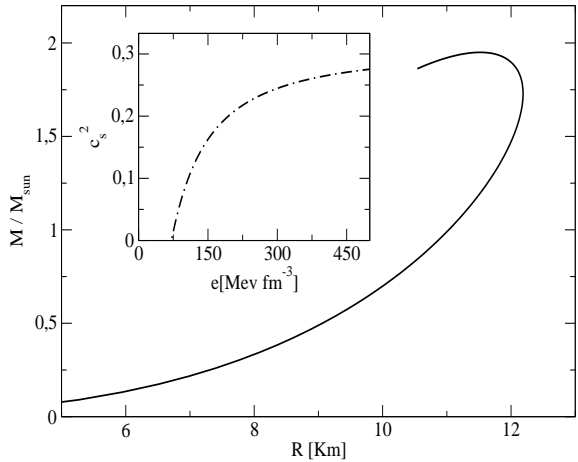}
\caption{Mass-Radius relation corresponding to the CCDM equation of state. In the insert, we display the speed of sound as a function of the energy density.}
\end{centering}
\end{figure}

We show in Fig.1 and Fig.2 the strange quark mass and the energy per
baryon for typical sets of parameters used in these models (see
details in a forthcoming paper \cite{nuovo}).  The parameter fixing is performed,
as in \cite{Zacchi}, by reproducing the tree level mass spectrum of
pseudoscalar mesons. From Fig.1 one can notice that the strange quark
mass decreases very slowly with density within the CM and SCM while it
presents a significant drop, within the CCDM, already at a density
comparable to nuclear saturation density. Concerning the minimum of
the energy per baryon, it is not possible to find a parameter set
allowing for the stability of strange quark matter for CM and SCM (in
agreement with previous findings \cite{Klahn:2015mfa,Wang}); in Fig.2
we show only one typical example for each model: within the CM a
minimum of the energy per baryon could be found but its value is
larger than the one of nuclear matter. Within the SCM, no minimum of
the energy per baryon is found unless in some specific fine tuned sets
of parameters. \footnote{Both in CM and SCM it is possible to find
  minima of the energy per baryon smaller than the one of nuclear
  matter but without a net strangeness fraction. Moreover, 
chiral symmetry restoration in the up and down quark sectors occurs
  at too low densities. Such cases are clearly ruled out.}. On the
other hand, within the CCDM, it is possible to find sets of parameters
for which the Bodmer-Witten hypothesis is fulfilled. In the case shown
in Fig.2, the minimum of the energy per baryon is $E/A \sim 913$ MeV
and it is realized at a density slightly larger than the nuclear saturation
density. This result is explained by the fast drop of the strange
quark mass, see Fig.1, which in turn is related to a density dependence of the masses 
that is introduced by the confinement dynamics.
An important constraint on quark matter
equation of state stems 
from the maximum mass of compact stars.
To reach values of maximum masses close to $2M_{\odot}$,
one must introduce within the CCDM also vector meson contributions.
We use here a quark-vector meson interaction term analogous to the
quark-scalar one, namely:
\begin{equation}
  \mathcal{L}_{int} =\,
  -\frac{\sqrt{2} g_{\sigma}}{\chi}\, (\bar{q}Mq) - \frac{\sqrt{2}
    g_{\omega}}{\chi}\, (q^{\dagger} V_{0} q)
\end{equation}
where  $M\,,V_0$ are respectively the scalar and vector meson matrices in the mean
field approximation. To our knowledge, no attempt to introduce the
vector mesons in the CCDM is present in the literature.
The quark star mass-radius relation as well as the speed of sound in
dense quark matter, obtained within the CCDM model, are displayed in
Fig.3.  The maximum mass is of $\sim 1.94 M_{\odot}$ and thus
compatible (within error bars) with the astrophysical measurements
\cite{Demorest:2010bx,Antoniadis:2013pzd}. We expect that a
calculation that introduces also the Fock term in the field equations
would provide larger maximum masses.  Finally, we would like to remark
that the results obtained within the CCDM model and here presented
correspond to a choice of the parameter set which is compatible with
parameter sets used in the literature for studying hadronic physics
observables, see \cite{Drago:1995ah} and Refs. therein.

\section{Maximum mass and radii of strange quark stars}
\label{sec-2}
We discuss now which are the astrophysical observations that could
test the Bodmer-Witten hypothesis. For the sake of simplicity
we use here a parametrization of the quark matter equation of
state in which the speed of sound $c_s$ is constant. The
relation between pressure and energy density is thus given by $p=c_s^2
(e-e_0)$ where $e_0$ represents the energy density at zero pressure
(i.e. the energy density of a droplet of quark matter). We note that
this equation of state has been also used in
Refs.\cite{Lattimer:2010uk} to investigate maximally compact stellar
configurations and in Ref.\cite{Alford:2015dpa} to construct hybrid stars. By indicating with $n_0$ the baryon density
corresponding to $e_0$
one can determine the following relation between pressure and
$n$ by simply using the thermodynamic relation $p=n^2\frac{\partial/n}{\partial n}$:

\begin{equation}
 p=k((n/n_0)^{1+c_s^2}-1)
\end{equation}

where $k=\frac{e_0 c_s^2}{1+c_s^2}$, see also \cite{Zdunik:2000xx}. Different sets of the three
parameters $c_s$,$e_0$ and $n_0$ determine different possible
equations of state with the corresponding mass-radius curves. In many
calculations of the quark matter equation of state it results that
$c_s^2 \lesssim 1/3$ (see \cite{Alford:2015dpa,Bedaque:2014sqa}) and we will therefore keep it fixed to $1/3$. The
values of $n_0$ and $e_0$ are on the other hand completely unknown. A
few constraints can however be used: $830$ MeV $\lesssim e_0/n_0=E/A < 930$
MeV if strange quark matter is assumed to be absolutely stable (the Bodmer-Witten
hypothesis) while preserving the stability of Iron with respect to two-flavor quark matter \cite{Witten:1984rs}. 
Moreover, we impose that $n_0 \gtrsim n_{sat}$ i.e. the baryon density of strange quark matter droplets must not be smaller than
the nuclear matter saturation density.
This constraint is satisfied both in the MIT bag
model and the CCDM once the parameters are fixed to reproduce basic properties of hadrons such as the radii \cite{Drago:1995ah}.

\begin{figure}[ptb]
\vskip 0.5cm
\begin{centering}
\epsfig{file=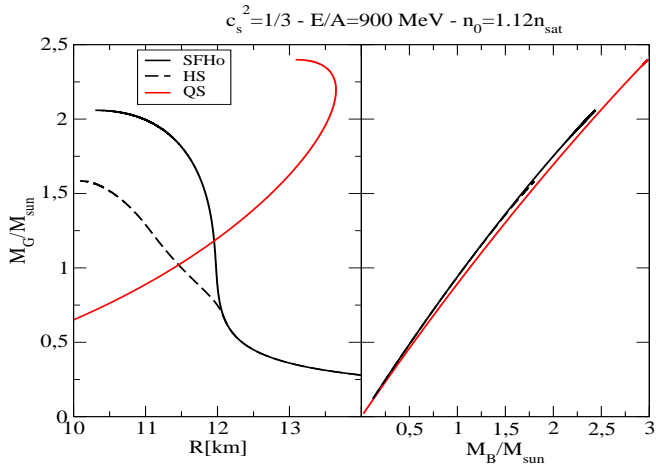,height=8.5cm,width=6cm,angle=-90}
\caption{(Left panel): mass radius plot for for nucleonic stars, hadronic stars (as obtained by using the SFHo equation of state \cite{Steiner:2012rk})
  and strange quark stars (as obtained by using a constant-speed equation of state with parameters tuned in order to achieve a maximum mass
of $2.4M_{\odot}$). In the right panel, the corresponding relations between baryonic mass and gravitational mass are displayed.}
\end{centering}
\end{figure}

In Fig.4, left panel, we show an example of quark star mass-radius
curve obtained by requiring that the maximum mass is of $2.4M_{\odot}$
(which is the value inferred for the black widow pulsar
\cite{vanKerkwijk:2010mt}) and that $e_0/n_0=900$, i.e. only 30 MeV
smaller than the binding energy of nuclear matter. The maximum mass
constraint allows to fix $n_0=1.12n_{sat}$. Notice that it is possible
to achieve such high values of the maximum mass in quark models provided
that $n_0$ is close to nuclear saturation density. 
For comparison we display
also the mass-radius relation for the nucleonic equation of state SFHo
\cite{Steiner:2012rk} without (solid line) and with hyperons and deltas (dashed line), \cite{Drago:2014oja}.
It is remarkable that while quark stars (with
masses above $1M_{\odot}$) have radii larger than nucleonic and hadronic stars,
they are more bound than them: by looking at the relation
between baryonic mass and gravitational mass (right panel of Fig.4)
one notices that at fixed baryonic mass strange quark stars are lighter than
nucleonic and hadronic stars, see also Ref.\cite{Drago:2013fsa}. It is clear therefore
that a process of conversion of a hadronic star into a quark star is
energetically convenient even if the radius of the final configuration
is larger than the radius of the initial configuration. Notice that
the gravitational mass of a strange star is smaller than the one of a hadronic star
(with the same baryonic mass) just because the Witten hypothesis is satisfied, almost
independently from the parameters' values.
The scenario
of the conversion of hadronic stars into strange quark stars has been
developed in several papers, see
Refs. \cite{Drago:2013fsa,Drago:2015fpa,Pagliara:2013tza,Drago:2015cea,Drago:2015dea}
and the phenomenological connections with gamma-ray-bursts
have been studied in Refs.~\cite{Drago:2015qwa,Pili:2016hqo}.
It is important to stress that in this scenario the two families of
stars, hadronic stars and quark stars, do coexist and that the
transition between the two branches is provided by the appearance, in
a hadronic star, of hyperons: once a sizable fraction of strangeness
is present through hyperons, the (metastable) stellar system can
convert into a pure strange quark star. The appearance of hyperons, as
obtained by many calculations, should take place around 2-3 times
saturation density, which in turn implies that only hadronic stars
more massive than about $1.5-1.6 M_{\odot}$ can convert, see
\cite{Drago:2015cea,Drago:2015dea}.

Presently, the highest measured
mass is the one of PSR J0348+0432, $M=2.01 \pm 0.04M_{\odot}$ \cite{Antoniadis:2013pzd}, but a candidate with a higher
mass does exist: the black widow pulsar with an estimated mass of
$2.4M_{\odot}$ \cite{vanKerkwijk:2010mt}. Moreover, the observations of internal plateaux in short gamma-ray-bursts originating from the
merger of two neutron stars seem also to suggest the existence of
stars with masses significantly larger than $2M_{\odot}$
\cite{Lu:2015rta,Li:2016khf}. Strange quark stars could have maximum masses
larger than $2M_{\odot}$ (see Fig.4-5) provided that the
baryon density of strange quark matter droplets is close enough to
the density of saturation of nuclei, see also \cite{Lattimer:2010uk}. This can be seen in Fig.5 in which we display the lines of
constant maximum mass in the plane of the two parameters $n_0/n_{sat}$
and $E/A$. 
Notice that, on the other hand, it is very difficult to obtain hadronic stars with masses
of $2M_{\odot}$ or above because of the softening of the equation of state
related to the appearance of hyperons and deltas. In this respect,
measurements of masses larger than $2M_{\odot}$ would favor the strange quark star
scenario.
We remark that this scenario is not in contradiction with
neutron matter calculations and terrestrial laboratory
experiments. Indeed the conversion of metastable hadronic stars could
be triggered only when the central density of such stars is high
enough that a significant amount of strangeness appears.

Let us now discuss radii measurements. Future experiments, such as
NICER, will measure the radius of a few compact stars with an error of
about 1km. As discusses in \cite{Drago:2015cea} and as one can see in
Fig.4, it is clear that if $R_{1.4} \lesssim 11$ km the scenario of
the two families of compact stars, and therefore the existence of
strange quark stars, will be proven \footnote{Notice that the model we
  have adopted for the nucleonic equation of state, the SFHo model,
  predicts a radius $R_{1.4}$ of about 12 km. This value is compatible
  with ab-initio calculations of neutron matter as obtained within the
  chiral effective field theory approach and quantum Monte-Carlo
  simulations \cite{Lattimer:2015nhk}.}.

\begin{figure}[ptb]
\vskip 1.5cm
\begin{centering}
\epsfig{file=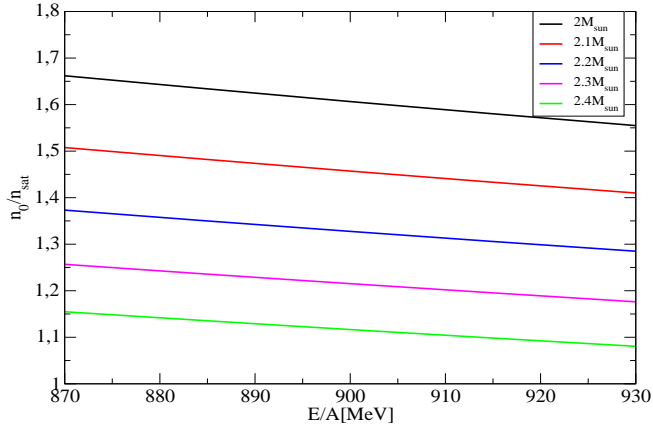,height=8.5cm,width=5.5cm,angle=-90}
\caption{Lines of constant maximum mass as function of the parameters $E/A$ and $n_0/n_{sat}$.}
\end{centering}
\end{figure}

\section{Conclusions}
We have discussed the possibility of the existence of strange quark matter
first within a purely theoretical approach, by using three different chiral models for quark matter
and after within phenomenological approach through
the measurements of masses and radii of
compact stars. The Bodmer-Witten hypothesis could be fulfilled only if spontaneous chiral symmetry
breaking and dynamical confinement are both implemented in the adopted quark model.
In particular we have shown that a pure SU(3) quark meson chiral model, even in the presence of
the additional dilaton field, does not allow for the existence of strange quark matter.
Instead,  in a simple confining chiral model, the chiral chromo-dielectric model, it is
possible to find a window of parameters for which strange quark matter
is absolutely stable. It is remarkable that the sets of parameters corresponding
to these cases, are quite similar to the sets of parameters usually adopted
for the study of hadronic observables.

Finally, we have discussed what are the implications for the equation of state
from the existence of massive compact stars and how future measurements of radii can be used
to test fundamental properties of dense matter.
In particular, a strong indication of the validity of the Bodmer-Witten
hypothesis will be measurements of $R_{1.4} \lesssim 11$ km. 
The existence of stars as massive as  $2M_{\odot}$ (or more)
would indicate that if the Bodmer-Witten hypothesis is correct, the (second)
minimum of the energy per baryon of strongly interacting matter is located
at a density slightly above the nuclear matter saturation density.
Differently from the minimum corresponding to nuclear matter,
this second minimum would contain a net strangeness fraction close to 1/3.
The possibility to search for this kind of matter also in terrestrial experiments
would be in this case very promising.

\end{document}